\documentclass[aps,prl,twocolumn,showpacs,superscriptaddress,groupedaddress]{revtex4}

\usepackage[utf8x]{inputenc}
\usepackage{amsmath,amssymb,amsfonts,subfigure,braket,color,dsfont}
\usepackage[english]{babel}
\usepackage{graphicx} 
\usepackage{sistyle}
\usepackage{listings}
\usepackage{gensymb} 

\begin{document} 
 
\title{Mechanical tunability of an ultra-narrow spectral feature with uniaxial stress}
\author{N. Galland} 
\affiliation{Univ. Grenoble Alpes, CNRS, Grenoble INP and Institut N\' eel, 38000 Grenoble, France}
\affiliation{LNE-SYRTE, Observatoire de Paris, Universit\' e PSL, CNRS, Sorbonne Universit\' e, Paris, France}
\author{N. Lu{\v c}i\'c} 
\affiliation{LNE-SYRTE, Observatoire de Paris, Universit\' e PSL, CNRS, Sorbonne Universit\' e, Paris, France}
\author{B. Fang} 
\affiliation{LNE-SYRTE, Observatoire de Paris, Universit\' e PSL, CNRS, Sorbonne Universit\' e, Paris, France}
\author{S. Zhang} 
\affiliation{LNE-SYRTE, Observatoire de Paris, Universit\' e PSL, CNRS, Sorbonne Universit\' e, Paris, France}
\author{R. Le Targat} 
\affiliation{LNE-SYRTE, Observatoire de Paris, Universit\' e PSL, CNRS, Sorbonne Universit\' e, Paris, France}
\author{A. Ferrier} 
\affiliation{Chimie ParisTech, Universit\' e PSL, CNRS, Institut de Recherche de Chimie Paris, 75005 Paris, France} 
\affiliation{Sorbonne Universit\'e, Facult\'e des Sciences et Ing\'enierie, UFR 933, 75005 Paris, France} 
\author{P. Goldner} 
\affiliation{Chimie ParisTech, Universit\' e PSL, CNRS, Institut de Recherche de Chimie Paris, 75005 Paris, France} 
\author{S. Seidelin}\email{signe.seidelin@neel.cnrs.fr}
\affiliation{Univ. Grenoble Alpes, CNRS, Grenoble INP and Institut N\' eel, 38000 Grenoble, France}
\affiliation{Institut Universitaire de France, 103 Boulevard Saint-Michel, F-75005 Paris, France}
\author{Y. Le Coq}
\affiliation{LNE-SYRTE, Observatoire de Paris, Universit\' e PSL, CNRS, Sorbonne Universit\' e, Paris, France}

\date{\today}

\begin{abstract}

Rare-earth doped crystals have numerous applications ranging from frequency metrology to quantum information processing. To fully benefit from their exceptional coherence properties, the effect of mechanical strain on the energy levels of the dopants - whether it is a resource or perturbation - needs to be considered. We demonstrate that by applying uniaxial stress to a rare-earth doped crystal containing a spectral hole, we can shift the hole by a controlled amount that is larger than the width of the hole.  We deduce the sensitivity of $\rm Eu^{3+}$ ions in an $\rm Y_2SiO_5$ matrix as a function of crystal site and the crystalline axis along which the stress is applied.

\end{abstract}


\maketitle

Rare-earth ions embedded in a crystalline matrix, at cryogenic temperatures, exhibit optical transitions with excellent coherence properties~\cite{Thiel2011} combined with the ease of use of solid state materials. Such properties can  be used for example in classical~\cite{Berger2016} and quantum~\cite{Nilson2005,Bussieres2014,Walther2015,Maring2017} information processing schemes, quantum optical memories~\cite{Zhong2017,Laplane2017}, quantum probes of photonic effects~\cite{Tielrooij2015}, and in ultra-high-precision laser stabilisation and spectroscopy~\cite{Julsgaard2007,Thorpe2011,Gobron2017}. In these materials, randomly distributed perturbations from the local matrix result in a broad inhomogeneous profile of the ion absorption spectrum, but spectral hole burning techniques can be used to realize narrow spectral features with a resolution only limited by the individual doping ions. Moreover, as the spectral properties of the individual ions are sensitive to the surrounding crystalline matrix, external stress applied to the crystal in the plastic regime will frequency displace a previously imprinted spectral hole, providing an interesting resource for tuning spectral holes reversibly, in addition to probing stress fields in a spatially resolved manner. Other applications of stress sensitivity can be found in the field of quantum opto-mechanics~\cite{Aspelmeyer2012}, where the vibrations of a mechanical resonator modulate the energy levels of quantum two-level system emitters embedded in the resonator material itself, as observed experimentally in resonators containing quantum dots~\cite{Yeo2014} or Nitrogen-Vacancy centers~\cite{Teissier2014, Ovartchaiyapong2014,Macquarrie2017}, and proposed theoretically for rare-earth ions doped resonators~\cite{Molmer2016,seidelin2019}.

In this work, we concentrate on $\rm Eu^{3+}$ ions in a $\rm Y_2SiO_5$ host matrix (Eu:YSO) at a temperature of 3.15\,K, as this material exhibits one of the narrowest optical transitions among solid-state emitters. We exploit the 580\,nm optical transition $^7F_0 \rightarrow$  $^5D_0$ which possesses near lifetime limited coherence times in the ms range~\cite{Equall1994,Oswald2018}. The YSO crystal has two non-equivalent locations within the unit cell where $\rm Eu^{3+}$ can substitute for $\rm Y^{3+}$, referred to as site 1 and 2 (vacuum wavelengths of 580.04\,nm and 580.21\,nm, respectively). We use the technique of spectral hole burning to benefit from the narrow homogeneous linewidth and at the same time the large signal to noise ratio coming from working with an ensemble of ions. Spectral holes are formed by resonant optical excitation and decay of the ions into other long-lived (several hours) dark states, in our case other hyperfine states within the electronic ground state manifold~\cite{Konz2003}. Subsequently scanning the frequency of a weak optical field across the hole will allow observing the structure. The impact of the scanning on the long-lived persistent spectral features is negligible during the typical time constant of the experiment (a few minutes for 5 to 6 scans).
  
\begin{figure}[t]
\centering
\includegraphics[width=80mm]{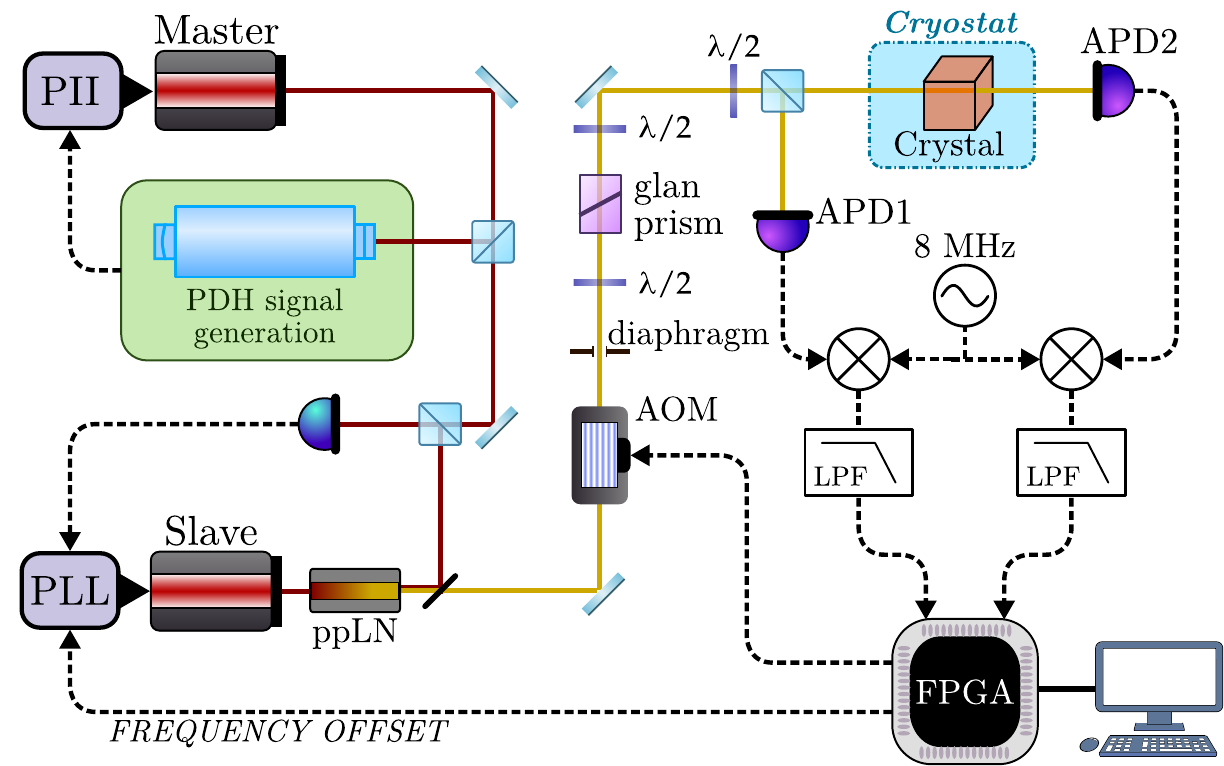}
\caption{\label{schema} Schematics of the experimental setup used for measuring the strain-sensitivity. The following abbreviations are used: ML Master Laser; SL Slave Laser; PDH Pound Drever Hall; APD1 and APD2 Avalanche Photo Detector 1 and 2; PLL Phase Lock Loop; PII Proportional double Integrator corrector; AOM Acusto-Optical Modulator; FPGA Field-Programmable Gate Array and LPF Low Pass Filter. The $\lambda$/2 waveplates and Glan polarizer are used to adjust the intensity of the optical power.}
\end{figure}


The experimental setup is shown in fig.~\ref{schema}. The laser system is based on two fiber-coupled extended cavity cw diode lasers (ECDL), referred to as master and slave laser, operating at 1160\,nm, delivering 65\,mW each. The master laser is frequency locked (locking bandwidth around 800\,kHz) by the Pound-Drever-Hall method to a commercial reference cavity exhibiting a fractional frequency instability below $10^{-14}$ for time constants of 1-100\,s, producing a radiation with a few Hz linewidth. The slave laser is offset phase locked to the master laser (lock bandwidth of typically 1\,MHz) and frequency doubled in a PPLN waveguide (with  free space output) to produce 4.5\,mW of radiation at 580\,nm. The slave laser at 580\,nm benefits from the spectral purity and stability imposed by the reference cavity, but with an offset frequency that can be tuned.

Further control of the 580\,nm laser field generated by the slave laser is provided by an acousto-optic modulator (AOM) used in double-pass configuration and driven by an RF signal generated by a computer controlled FPGA (Field-Programmable Gate Array). By choosing accordingly the reference Fabry-Perot cavity mode onto which the master laser is frequency locked, and the frequency difference between the master and the slave lasers, we ensure that the frequency of the slave laser at 580\,nm is positioned near the center of the Eu:YSO crystal absorption profile (for either site 1 or site 2). The 580\,nm light beam from the slave laser then passes through a $\lambda$/2 waveplate and a Glan polarizer for tunable attenuation, and lenses are used to adjust the beam diameter to 6\,mm, followed by a tunable apperture which allows restricting the beamsize further. A second polarizing beam splitter then allows splitting the beam in two parts, one sent directly to a silicon avalanche photodiode (APD1 in fig.~\ref{schema}), the other passing through the crystal after polarization tuning via a final $\lambda$/2 waveplate, before being sent to a second avalanche photodiode (APD2). This last waveplate allows to tune the polarization (parallel to the D$_1$-axis) for optimum absorption in the crystal~\cite{Ferrier2016}.

When burning a spectral hole, the first waveplate is rotated such that the optical power is maximum at the crystal and a single tone is applied to the AOM, producing typically 70\,$\mu$W/cm$^{2}$ during 0.5\,s. When probing the hole, the AOM is driven with an appropriate RF waveform so as to produce two optical tones~\cite{Jobez2016}, separated by 800\,kHz, jointly frequency scanned at 100\,kHz/s, one of them in the vicinity of the spectral hole. The first waveplate is also rotated to decrease optical power (400\,nW/cm$^{2}$ in each mode). The corresponding detected beatnotes on the APDs (800\,kHz) are up-converted to 8.8\,MHz (by mixing with an 8\,MHz signal), digitized, digitally down-converted to base-band and streamed to a computer, which filters them in a 15\,kHz bandwidth and extracts the transmission of the crystal from the comparison between the amplitude of the signals from APD1 and APD2. This particular data processing allows minimizing excess low-frequency noise of photodiodes, amplifiers and analog to digital converters.

We use a commercial closed cooling cycle cryostat (OptiDry from MyCryoFirm) to maintain the crystal at 3.15\,K with a  passive vibration isolation stage between the cooler and the science chamber and a commercial active vibration canceling platform supporting the science chamber itself. The chamber is composed of a thermally stabilized 3.15\,K copper plate and enclosing shield, with 25\,mm diameter windows for optical access.

\begin{figure}[t]
\centering
\includegraphics[scale=1]{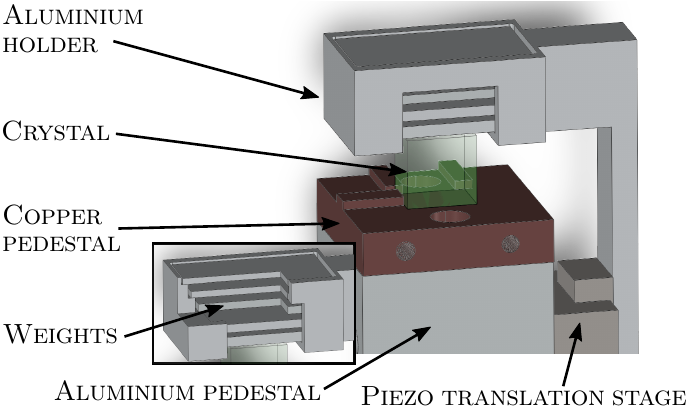}
\caption{\label{holder} Schematics of mount and motorized piezo-stage, with which we can deposit (or remove) a given number of weights on the top of the crystal inside the cryostat, allowing to apply several different values of a calibrated stress to the crystal. The inset shows a zoom on the inside of the staircase-like structure.}
\end{figure}

The Eu:YSO crystal is grown by Czochralski process with 0.1 at.\% europium doping. After x-ray orientation, it is cut into a $8\times8\times4\!$\,mm$^3$ cuboid, such that the optical beam can propagate along the crystallographic $b$-axis, and the two other facets are oriented along the principal dielectric axes D$_1$ and D$_2$, parallel to which we wish to apply mechanical stress. The two largest facets, perpendicular to the $b$-axis, are polished for optical beam propagation and set with a small relative angle (2\degree) to prevent intrusive Fabry-Perot cavity build-up. In order to apply the stress parallel to the D$_1$-axis, the crystal is oriented such that this axis is in the vertical direction. We then rotate the crystal 90\degree  to perform the measurements with a stress applied parallel to the D$_2$-axis. The crystal is mounted in the science chamber inside the cryostat, on a specially designed pedestal, see fig.~\ref{holder}.

\begin{figure*}[t] 
\includegraphics[scale=1]{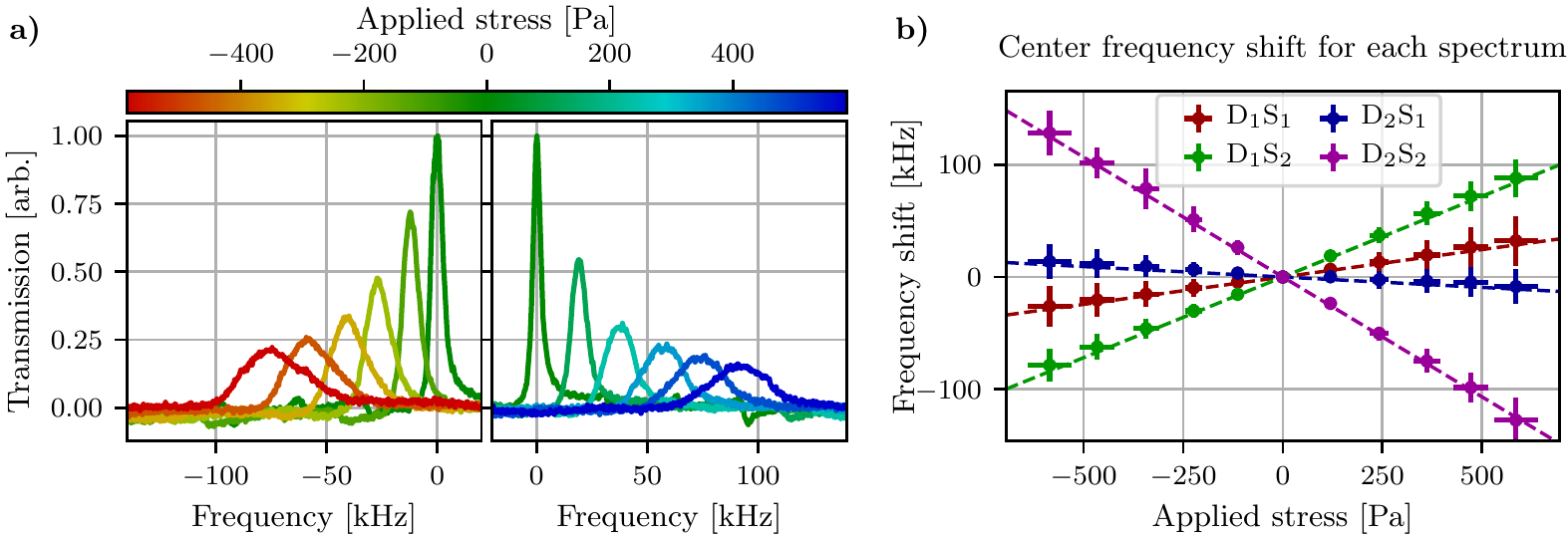}
\caption{\label{D1S2}  When stress is applied to the crystal, the profiles of the spectral holes shifts in frequency. a) Behavior of spectral hole under stress corresponding to the crystal site 2, with stress applied parallel to D$_1$ (D$_1$S$_2$). Positive values of the stress correspond to adding weights one by one following the spectral hole burning, whereas negative values correspond to burning the holes with all the weights on the top of the crystal, and then, removing them, one by one. b) Displacements of the center of the spectral hole for different axes (D$_1$ and D$_2$) parallel to which the stress is applied and different crystal sites (S$_1$ and S$_2$), with linear fits to the data.}
\end{figure*}     
The strain-sensitivity of the $\rm Eu^{3+}$ ions in the YSO matrix when applying an isotropic pressure by changing the helium gas pressure inside a cryostat has been measured by another group~\cite{Thorpe2011}. However, for all applications discussed above, the uniaxial strain sensitivity is needed, which requires applying a calibrated force inside the cryostat in a single, given direction. To achieve this, we have used a staircase-like structure (see inset in fig.~\ref{holder}) of which the vertical position is controlled by a motorized platform situated inside the cryostat, providing a simple way of placing a variable number of objects on the top of the crystal. The weights of the objects have been measured prior to insertion in the cryostat in order to provide a well-calibrated force. We start the sequence by burning a spectral hole with no weights on the crystal, and we then add them one by one, each time recording the profile of the spectral hole. The beam diameter for probing is approx. 2\,mm (which is small compared to the polished facet of the crystal). For verification, the measurements are also repeated by burning the spectral hole with all weights on the top of the crystal, and then removing them one by one, recording the spectra (corresponding to negative stress). The measurements are also repeated in several different locations of the crystal, to ensure that the strain field is homogeneous.

To create a sufficiently homogeneous strain field, we use a 3-axis piezo translation stage, 2 of these axes being used for fine centering of the weights on the crystal, and the last, vertical axis, for operating the staircase deposition structure. As the energy transitions in $\rm Eu^{3+}$ are not only sensitive to mechanical strain, but also electric fields~\cite{Thorpe2013, Mcfarlane2014,Li2016}, special care is required to eliminate uncontrolled residual electrical field and charges. In particular, the motorized platform and crystal both have been surrounded by independently grounded aluminum shields and cover. Finally, for the weights to produce homogeneous strain in the crystal, we found it necessary to add a layer (approx. 1\,mm thick) of powderized YSO which does not contain europium ions (micrometer typical grain size) on the top and bottom of the crystal. We further noticed that the silver lacquer commonly used in cryogenics for ensuring good thermal contact between the crystal and the cold finger was producing comparatively large, unpredictable stress to the crystal and had to be avoided. The thermalization of the crystal relying solely on radiation and heat-exchange through the top and bottom powder layers, a typical waiting time of 5 minutes is used to reach steady state when necessary, in particular after operation of the translation stages for relatively large (more than 5\,mm) excursions (which was found to momentarily perturb the temperature stability). The frequency drift of the cavity is continuously monitored using the beatnote of our laser with a frequency comb (not shown in fig.~\ref{schema}), but the drift consistently turns out to be negligible.

An example of the displacement of a spectral hole consisting of ions occupying the crystal site 2 when applying uniaxial stress  along the D$_1$ crystalline axis is shown in fig.~\ref{D1S2} a). In addition to the displacement of the center of the spectral hole, we also observe a broadening of the profile, which is partly due to inhomogeneities on the microscopic level in the different ions' response to applied strain, as observed in other rare-earth doped systems~\cite{Reeves1989,Chauvet2019}, partly due to a residual inhomogeneity in the applied strain.

The measurements are repeated when applying stress along both the D$_1$ and D$_2$ crystalline axis and for both crystal site 1 and 2. In fig.~\ref{D1S2} b) we plot the frequency shift of the center of the spectral hole as a function of the stress applied to the crystal, for the different axes and crystal sites. The largest contribution to the errors stems from the inhomogeneity in direction and amplitude of the strain. We include this uncertainty by assigning horizontal error-bars inferred from measurements of the displacements in different locations in the crystal (plus a minor contribution from the uncertainty of the measurements of the weights), and a vertical error-bar which reflects the full-width-half-maximum of the spectral hole.


We observe a systematic tendency of some data points to be marginally above the linear fit in fig.~\ref{D1S2} b). This phenomenon may need a further study, but a possible explanation is that the operation of the translation stage locally creates a minute increase in temperature (despite the waiting time) at the position of the crystal. As the resonant frequency of the hole is also dependent on the temperature~\cite{Thorpe2011}, the heating gives rise to a small systematic frequency shift which is positive for both positive and negative stresses applied, and has no or negligible impact on the final values of the slopes extracted from the linear fits. Finally, we have repeated the measurements for spectral holes situated in different parts of the inhomogeneously broadened profile, and again, we obtain stress sensitivities confined to within the error-bars.
We observe that, within the range of the measurement, the central shifts are linear with the applied stress and extract proportionality coefficients from fits. The weighted average of several data sets provides the final values, shown in the table below, where the stated uncertainty stems from a combination of the square root of the relevant diagonal elements of the covariance matrix terms for each fit and repeatability of the measurement series:
\vspace{0.1 cm}
\begin{center}
\begin{tabular}{ccrclcrcl}
		\hline
		Axis & ~Site & \multicolumn{3}{c}{Hz Pa$^{-1}$} & ~ & \multicolumn{3}{c}{THz Strain$^{-1}$}  \\
		\hline\hline
		D1 & 1 & 46  & $\pm$ & 17~~ & & 6.2   & $\pm$ & 2.3  \\
		D1 & 2 & 137 & $\pm$ & 16  & & 18  & $\pm$ & 2.2  \\
		\hline
		D2 & 1 & -19 & $\pm$ & 10  & & -2.6  & $\pm$ &  1.4  \\
		D2 & 2 & -213 & $\pm$ & 13 & & -29 & $\pm$ &  1.8 \\
		
		\hline
\end{tabular}
\end{center}

\vspace{0.1 cm}

As described above, these values are obtained by studying the displacement of a spectral hole, but the values directly apply to the shifts of the energy levels of the individual europium ions constituting the spectral hole. 

Other systems potentially suitable for strain-coupled quantum systems such as NV and SiV centers in diamond tend to exhibit a somewhat higher sensitivity for optical transitions. In particular, maximum sensitivities as high as 1000\,Hz/Pa in NV$^{-}$ centers~\cite{Davies1976} and 1800\,Hz/Pa in SiV$^{-}$ centers~\cite{Meesala2018} have been reported, a factor of five to ten larger than for the system reported here. However, the sensitivity should be compared to the linewidth of the corresponding optical transitions, which can be as low as 122 Hz for the considered transition in Eu:YSO~\cite{Equall1994}, whereas the corresponding optical transitions (for the zero-phonon line in bulk material) is 13\,MHz in NV$^{-}$~\cite{Tamarat2006} and 322\,MHz in SiV$^{-}$~\cite{Evans2016}, both already close to the lifetime limited value. This is at least 5 orders of magnitude larger than for the europium ions.

Although for most physical applications, the sensitivity to stress is the most relevant quantity, in order to compare different systems from a microscopic point of view, the sensitivity per strain is preferably used, which is independent of the stiffness of the crystal. These values, also given in the table, are calculated using a Young modulus for YSO of 135\,GPa. Due to the high Young modulus in diamond (1220\,GPa), on a microscopic level, the strain sensitivity is increased by an additional order of magnitude in these systems. The smaller sensitivity for the rare-earth dopants partly stems from the protection of the $4f$-orbitals (within which the optical transition takes place) by the outer $s$ and $p$ orbitals - the same effect being in part the origin of the long coherence times. 

The strain sensitivity in Eu:YSO can be used in a variety of contexts. Utilized as a differential accelerometer or relative gravimeter, with a test mass of 1\,kg and staying well below the elastic limit of YSO (approx. 100\,MPa~\cite{Sun2008}), our $8\times8\times4\!$~mm$^3$ crystal in this configuration would exhibit a frequency shift of 6.6\,MHz per m.s$^{-2}$ in the most sensitive site and direction. Assuming shot noise limited dispersion-based detection of the position of the spectral hole~\cite{Gobron2017} with typical slope of 0.1\,mrad/Hz, with 100\,nW of probing power, this corresponds to a sensitivity of $1\times10^{-6}$ rad/$\sqrt{\rm{Hz}}$ / (660 rad/m.s$^{-2}$) $=1.4\times$10$^{-9}$ m.s$^{-2}/\sqrt(\rm{Hz})$. This is competitive with all existing commercial technology accelerometers, and could hold promise as an interesting novel application of rare-earth doped materials. 

On the other hand, in applications requiring extreme frequency stability of the spectral hole, such as ultra-stable cw laser realization for optical atomic clocks~\cite{Thorpe2011,Gobron2017}, the sensitivity sets a limit to the level of acceptable residual mechanical vibrations. For instance, in order to obtain a laser with a relative frequency stability of 10$^{-18}$ at 1\,s by locking it to a spectral hole would require, considering only the least sensitive direction and crystal site, a noise level below approx. 10$^{-6}$\,m.s$^{-2}/\sqrt(\rm{Hz})$ near 1\,Hz frequency  (which is likely within reach). Special mounting of the crystal, in the spirit of that described in ref.~\cite{Thorpe2011} or commonly used with ultra-stable Fabry-Perot cavities (see for example ref.~\cite{Millo2009}), can be designed, with the help of the presented measurements, to largely compensate the effect of acceleration-induced deformations due to residual vibrations. Finally, in the context of strain-coupled nano-optomechanics, the measured uniaxial strain sensitivity should be sufficiently large to provide a direct observation of quantum features of a mechanical resonator, such as the zero point motion~\cite{Molmer2016, seidelin2019}.

In conclusion, our demonstration of mechanical tunability of the frequency of a spectral hole in $\rm Eu^{3+}{:}Y_2SiO_5$ at cryogenic temperature, combined with quantitative measurements of the uniaxial stress sensitivity, evidence an important fundamental effect that may be utilized in a large variety of applications ranging from high precision measurements to quantum nano-optomechanics.  Moreover, due to the long lifetime of the spectral features, this tunable quantum hybrid system thus posseses an inherent memory, which is unique among solid-state emitters. The low degree of symmetry of the two non-equivalent substitution sites as well as of the crystal itself makes it extremely challenging to calculate the stress sensitivity coefficients {\it ab initio}. Therefore, our measurements are essential in order to predict the relevance and limitations of the many potential applications of this system, in addition to serving as an input for theoretical models. The strain sensitivity of the 580\,nm transition in $\rm Eu^{3+}{:}Y_2SiO_5$ is a factor of 5 to 10 times smaller than values measured for optical transitions in other dopants such as NV and SiV centers in diamond. However, the homogeneous linewidth for europium ions can be 5 orders of magnitude narrower than the corresponding optical transition in NV and SiV systems, making rare-earth doped systems interesting candidates for strain-engineered quantum systems relying on a transition which can be directly addressed optically.

\vspace{0.2 cm}
 
We thank A. C. Bleszynski Jayich, S. Kr\"{o}ll, S. Horvath, T. Chaneli\`ere, R. Ahlefeldt, K. M\o lmer, J. Sankey and E. Dupont-Ferrier for useful discussions. The project has been supported by the European Union's Horizon 2020 research and innovation program under grant agreement No 712721 (NanOQTech). It has also received support from the Ville de Paris Emergence Program, the R\'{e}gion Ile de France DIM C'nano and SIRTEQ, the LABEX Cluster of Excellence FIRST-TF (ANR-10-LABX-48-01) within the Program ``Investissements d'Avenir'' operated by the French National Research Agency (ANR), and the EMPIR 15SIB03 OC18 and from the EMPIR program co-financed by the Participating States and from the European Union's Horizon 2020 research and innovation program.

\end{document}